\documentclass[a4paper]{article}

\usepackage{graphicx}
\usepackage{color}

\newcommand{\MARKI}[1]{#1}
\newcommand{\MARKII}[1]{#1}
\newcommand{\MARKIII}[1]{#1}
\newcommand{\citep}[1]{\cite{#1}}
\newcommand{\citet}[1]{\cite{#1}}

\title{Electric solar wind sail applications overview}

\author{Pekka Janhunen, Petri Toivanen, Jouni Envall and Sini Merikallio\\
Finnish Meteorological Institute, Helsinki, Finland\\
\\
Giuditta Montesanti and Jose Gonzalez del Amo\\
European Space Agency, ESTEC, The Netherlands\\
\\
Urmas Kvell, Mart Noorma and Silver L\"att\\
Tartu Observatory, T\~oravere, Estonia
}

\begin{document}

\maketitle


\begin{abstract}
We analyse the potential of the electric solar wind sail for solar
system space missions. Applications studied include fly-by missions to
terrestrial planets (Venus, Mars and Phobos, Mercury) and asteroids,
missions based on non-Keplerian orbits \MARKI{(orbits that can be
  maintained only by applying 
  continuous propulsive force)}, one-way boosting to outer
solar system, off-Lagrange point space weather forecasting and
low-cost impactor probes for added science value to other missions. We
also discuss the generic idea of data clippers (returning large
volumes of high resolution scientific data from distant targets packed
in memory chips) and possible exploitation of asteroid
resources. \MARKI{Possible orbits were estimated} by orbit calculations assuming circular and
coplanar orbits for planets. Some particular challenge areas requiring
further research work and related to some more ambitious mission
scenarios are also identified and discussed.
\end{abstract}

\section{Introduction}

The electric solar wind sail (E-sail) is an advanced concept for
spacecraft propulsion, based on momentum transfer from the solar wind
plasma stream, intercepted by long and charged tethers
\citep{JanhunenAndSandroos2007}.  The electrostatic field created by
the tethers deflects trajectories of solar wind protons so that their
flow-aligned momentum component decreases.  The flow-aligned momentum
lost by the protons is transferred to the charged tether by a Coulomb
force (the charged tether is pulled by the plasma charge separation
electric field) and then transmitted to the spacecraft as thrust.  The
concept is attractive for applications because no propellant is needed
for travelling over long distances.  The E-sail's operating principle
is different from other propellantless propulsion technologies such as
the solar photon sail \citep{McInnes2004} and the solar wind magnetic
sail. The former is based on momentum transfer from sunlight (solar
photons) while the latter is based on a large loop-shaped
superconductive wire whose magnetic field deflects solar wind protons
from their originally straight trajectories
\citep{ZubrinAndAndrews1991}.

The main purpose of this article is to analyse the potential of E-sail
technology in some of the envisaged possible applications for solar
system space activities. To a limited extent we also adopt a
comparative approach, estimating the added value and other advantages
stemming from E-sail technology in comparison with present chemical
and electric propulsion systems and (in some cases) with other
propellantless propulsion concepts. When making such comparisons a key
quantity that we use for representing the mission cost is the total
required velocity change, \MARKI{$\Delta v$, also called delta-v}.

The Sail Propulsion Working Group, a joint working group between the
Navigation Guidance and Control Section and the Electric Propulsion
Section of the European Space Agency, has envisaged the study of three
reference missions which could be successfully carried out using
propellantless propulsion concepts. In particular, in the frame of the
Working Group, the following reference missions are studied: mission
to asteroids, mission to high inclination near-Sun orbits (for
solar polar observation) and a mission requiring non-Keplerian
orbits \MARKI{i.e.~orbits which can be maintained only by applying
  continuous propulsive thrust}. The possibility to apply the E-sail technology to these
missions is also discussed in this paper.

Currently, the demonstration mission ESTCube-1 is being developed by
Tartu University to provide a practical proof of the E-sail concept in
low Earth orbit (LEO) \citep{JanhunenEtAl2010}.  In LEO the E-sail
would sense a plasma stream moving at a relative velocity of $\sim 7$
km/s which is much slower than in the solar wind
(300-800 km/s), but at the same time with much higher plasma
density \MARKII{($\sim 10^{10}$--$10^{11}$ m$^{-3}$ versus $\sim 5\cdot 10^6$ m$^{-3}$
in the solar wind)}. After the LEO demonstration of ESTCube-1 it is important to
carry out a mission in the free-streaming solar wind outside Earth's
magnetosphere in order to fully demonstrate the feasibility of the
E-sail concept. One possibility for such solar wind test mission could
be a six unit (6U) CubeSat based on ESTCube-1 1U heritage, but launched
to geostationary transfer orbit (GTO) and then raised to a solar
wind intersecting orbit by an onboard butane cold gas thruster.

The related concept of electrostatic plasma brake for deorbiting a
satellite \citep{JanhunenPlasmaBrake} is only briefly touched upon in
section 3, otherwise it is left outside the scope of this paper.

\section{Mathematical \MARKI{background}}

The orbital calculations presented in this paper are based on assuming
circular and coplanar planetary orbits. For a semi-quantitative
discussion this approximation is sufficient, although some care is
needed with Mercury whose eccentricity is rather significant.

For computing impulsive propulsion \MARKI{$\Delta v$} values for orbital changes
we use the so-called vis-viva equation which is valid for elliptical
Keplerian orbits,
\begin{equation}
v = \sqrt{GM\left(\frac{2}{r}-\frac{2}{r_1+r_2}\right)}.
\label{eq:vis-viva}
\end{equation}
Here $G$ is Newton's constant of gravity, $M$ is the mass of the
central object, $r_1$ is the orbit's perigee distance, $r_2$ is the
apogee distance, $r$ is the instantaneous radial distance within the
orbit ($r_1 \le r \le r_2$) and $v$ is the orbital speed at $r$.

When an impulsive chemical burn is performed within the gravity well of a massive body, the orbital energy
changes more than in free space far from massive bodies. This so-called Oberth effect is central
to orbital dynamics and it is a direct consequence of the
conservation of energy. For example, assume that a spacecraft is in a parabolic
(marginally bound or marginally escaping) orbit around a planet and
that it makes an impulsive burn of magnitude \MARKI{$\Delta v$} at distance
$r$ from the planet's centre along its instantaneous orbital velocity
vector so that the spacecraft is ejected out from the planet with
hyperbolic excess speed $v$. If the spacecraft is in a parabolic (zero
total energy) orbit with respect to a central body of mass $M$, its
orbital speed at distance $r$ from the body is equal to the local escape speed
\begin{equation}
v_e = \sqrt{\frac{2 GM}{r}}
\end{equation}
as can also be deduced from Eq.~(\ref{eq:vis-viva}) by setting $r_2=\infty$.
After the impulsive burn we obtain by energy conservation
\begin{equation}
\frac{1}{2}v^2 = \frac{1}{2}\left(v_e+\Delta v\right)^2 - \frac{GM}{r}.
\end{equation}
Solving for the resulting hyperbolic excess speed $v$ we obtain
\begin{equation}
v = \sqrt{\left(\Delta v\right)^2 + 2 v_e \Delta v}.
\end{equation}

Conversely, if we need to find the magnitude of the burn \MARKI{$\Delta v$}
which yields a given hyperbolic excess speed $v$, the answer is
\begin{equation}
\Delta v = \sqrt{v^2 + v_e^2} - v_e.
\end{equation}
For example if a spacecraft is in parabolic Earth orbit, it needs a
heliocentric \MARKI{$\Delta v$} of 3 km/s to enter Mars transfer orbit. To
accomplish this requires a near-Earth burn of only 0.4 km/s
magnitude. For a heavy object such as Jupiter whose escape speed is 60
km/s the effect is even larger: a burn of 0.4 km/s made on parabolic
orbit near Jupiter gives a heliocentric \MARKI{$\Delta v$} of 7 km/s.

Low thrust transfer between circular orbits needs a \MARKI{$\Delta v$} which is
simply equal to the difference of the orbital speeds. Thus, slowly
spiralling out from low Earth orbit (LEO) to escape \MARKII{e.g.~by} electric
propulsion requires a total \MARKI{$\Delta v$} of 7.7 km/s although the same task
done with an impulsive near-Earth burn needs only 3.2
km/s. \MARKII{While proceeding via elliptical orbit would make the low
  thrust $\Delta v$ somewhat smaller than 7.7 km/s, doing so would
  necessitate turning the thruster off part of the time which might be
  wise in case of electric propulsion, but not necessarily wise in case
  of the E-sail where propellant consumption is not an issue.} Similar
considerations hold for a transfer between heliocentric orbits if the
transfer is slow. Thus, a low thrust propulsion system generally needs
to have a higher specific impulse than an impulsive method to provide
practical benefits. The electric sail fulfills this requirement nicely
because being a propellantless method, its specific impulse is infinite.

\MARKII{\section{E-sail performance and other properties}}

\label{sect:performance}

\MARKII{

Performance estimates of the E-sail are based on numerical simulation
\citep{JanhunenAndSandroos2007,Janhunen2009,Janhunen2012} and
semiempirical theory derived from them
\citep{Janhunen2009,Janhunen2010}. In laboratory conditions mimicing
LEO plasma ram flow, the electron sheath shape and width were measured
by \citet{SiguierEtAl2013} and one can calculate that the result is in good agreement with
theoretical formulas given by \citet{Janhunen2009}.

Mass budgets of E-sails of various thrust levels and for different
scientific payload masses were estimated and tabulated by
\citep{JanhunenEtAl2013}. For example, to yield characteristic
acceleration (E-sail thrust divided by total spacecraft mass at 1 au
distance from the sun in average solar wind) of 1 mm/s$^2$, the
spacecraft total mass was found to be 391 kg (including 20\% margin)
of which 143 kg is formed by the E-sail propulsion system consisting
of 44 tethers of 15.3 km length each. Characteristic acceleration of 1
mm/s$^2$ corresponds to 31.5 km/s of $\Delta v$ capability per year.

}

Outside \MARKII{of} Earth's magnetosphere, the E-sail can provide propulsive
thrust almost everywhere in the solar system. The only restrictions
are that the thrust direction cannot be changed by more than $\sim \pm
30^{\rm o}$ and that inside giant planet magnetospheres special
considerations are needed. The E-sail thrust magnitude decays as $\sim
1/r$ where $r$ is the solar distance. Notice that the E-sail thrust
decays slower than photonic sail and solar electric propulsion thrust
because the latter ones decay as $1/r^2$. \MARKII{The reason is that
  while the solar wind dynamic pressure decays as $1/r^2$, the
  effective area of the sail is proportional to the electron sheath
  width surrounding the tethers which scales similarly to the plasma
  Debye length which in the solar wind scales as $\sim r$. Since the
  E-sail thrust is proportional to the product of the dynamic pressure
  and the effective sail area, it scales as $1/r$.}

The solar wind is highly
variable and at first sight one might think that this would set
restrictions to applying the solar wind as a thrust source for space
missions. However, if the electron gun voltage is controlled in flight
so as to produce maximal thrust with available electric power, the
resulting E-sail thrust varies much less than the solar wind dynamic
pressure and accurate navigation is possible
\citep{ToivanenAndJanhunen2009}.

As mentioned above, by inclining the sail the thrust direction can be
modified by up to $\sim 30^{\rm o}$. This makes it possible to spiral
inward or outward in the solar system by tilting the sail in the
appropriate direction to decrease or increase the heliocentric orbital
speed, respectively. Thus, even though the radial component of the
E-sail thrust vector is always positive, one can still use the system
also to tack towards the sun. A similar tacking procedure is possible
also with photonic sails. Unlike most photon sails, E-sail thrust can
be throttled at will between zero and some maximum by controlling the
power of the electron gun.

\section{E-sail applications}

The number of potential E-sail applications is large. Here we use a
categorisation into five main groups: (1) asteroid and terrestrial
planets, (2) non-Keplerian orbits (e.g., off-Lagrange point solar wind
monitoring \MARKI{to achieve longer warning time for space weather
  forecasts}), (3) near-Sun missions, (4) one-way boosting to outer
solar system and (5) general ideas for impactors or penetrators,
``data clippers'' carrying data as payload and in situ resource
utilisation (ISRU).

The E-sail needs to be raised beyond Earth's magnetosphere before it
can generate propulsive thrust. The magnetosphere boundary
(magnetopause) position varies according to solar wind dynamic
pressure, but resides on average at $\sim$ 10 $R_E$ in the dayside and
much farther in the nightside (here $R_E$ is Earth's radius 6371.2
km). Lifting the orbital apogee to 20 $R_E$ requires 2.9 km/s
impulsive boost from low Earth orbit (LEO) assuming 300 km initial
altitude, while lifting to Moon distance (60 $R_E$) requires 3.1 km/s
and marginally escaping from the Earth-Moon system 3.2 km/s. One could
in principle start an E-sail mission from a 20 $R_E$ apogee
sun-oriented elliptical orbit, but it would necessitate some 1-2 km/s
of E-sail thrustings near the apogee to raise the nightside orbit from
LEO to 20 $R_E$ and beyond. Because the E-sail is about fast travel in
the solar system, one typically wants to start the mission
fast. Therefore in this paper we adopt 3.2 km/s (injection from LEO to
marginal Earth escape orbit) as the size of the impulsive chemical
burn that must be made at LEO to start an E-sail mission. The altitude
and inclination of the LEO orbit are not essential because once the
E-sail has reached the solar wind, it can correct its course. Also,
multiple E-sail probes targeted to different destinations can share
the same launcher to the initial escape orbit.

The E-sail needs the solar wind or other fast plasma stream to
work. Therefore the E-sail cannot in practice be used inside Earth's
magnetosphere where the plasma generally does not stream rapidly.  An
exception to this limitation is that one can use an E-sail like
apparatus for plasma braking in LEO \citep{JanhunenPlasmaBrake},
utilising the $\sim 7$ km/s speed difference between the orbiting
satellite and nearly stationary ionosphere and the fact that the
plasma density in LEO is high so that the process is relatively
efficient even though the speed difference is much less than the solar
wind speed of 400-800 km/s. In giant planet magnetospheres the plasma
corotates rapidly with the planet which might enable some form of
E-sailing also inside giant planet magnetospheres; this question
should be addressed in future studies. Mars and Venus have no
magnetospheres so that only some modifications to E-sail propulsion
are imposed by the existence of the plasma wake around those
planets. Mercury has a weak and small magnetosphere so that E-sailing
can be in general used, although not necessarily down to the lowest
orbits. We remark that the E-sail can obviously fly through
magnetospheres without limitations, only its ability to generate
propulsive thrust inside those regions is limited or absent, depending
on the case as described above.

When manoeuvring around planets, the capability of the E-sail tether
rig to tolerate eclipse periods should be analysed on a case by case
basis. Depending on the orbit's geometry and on the planet's
atmosphere and surface properties, the long E-sail tethers may
experience significant contraction due to rapid cooling if the
spacecraft flies into eclipse, and corresponding elongation takes
place when the eclipse period ends. If thermal contraction of the
tethers is rapid, it might cause harmful dynamical oscillations of the
tether rig. In this paper we do not take such possible restrictions
into account, but assume that E-sails can operate around planets
either by avoiding eclipses or by having the tether rig engineered so
as to tolerate the effects of thermal contraction.

\MARKII{As mentioned in Section \ref{sect:performance}, the
  E-sail performance has been estimated and tabulated earlier in
  various configurations \citep{JanhunenEtAl2013}. The obtainable performance
  (characteristic acceleration) depends on how large
  fraction the E-sail propulsion system forms of the spacecraft total
  mass. There is also a practical upper limit of E-sail size beyond which
  complexity would increase and performance would drop. At present
  level of technology this soft limit is likely to be $\sim 1$ N thrust at
  1 au solar distance. For small and moderate payloads up to few
  hundred kg mass, high performance is typically available, of order 1
  mm/s$^2$ characteristic acceleration corresponding to $\sim 30$ km/s
  of $\Delta v$ capability per year. The numbers must be scaled by $1/r$ when going
  beyond $1$ au.}

\MARKII{In the following subsections we survey various solar system
applications for the E-sail.} \MARKIII{In each case, we discuss how
  much chemical $\Delta v$ could be saved by using the E-sail. In
  some cases we also make comparisons with electric propulsion (ion
  engines, Hall thrusters and other low thrust electric rocket devices).}

\subsection{Terrestrial planets and asteroids}

\subsubsection{Venus}

Venus is the most nearby planet and the easiest to access in the
\MARKI{$\Delta v$} sense by impulsive kicks. An injection from marginal Earth
escape orbit to a Venus transfer orbit requires a 0.3 km/s near-Earth
kick and capturing from the Venus transfer orbit to high Venus orbit
requires 0.4 km/s near Venus. If the \MARKIII{target} is low Venus orbit, orbit
lowering needs additional 3.0 km/s or alternatively propellantless
gradual aerobraking. An E-sail could do any or all of these manoeuvres
with no propellant consumption. The transfer time is slightly longer
but comparable to impulsive kick Hohmann transfer
\citep{QuartaEtAl2010}. For example if the target is a low Venus orbit,
using the E-sail would save 0.7 km/s impulsive burns and make an
aerobraking device unnecessary, typically with comparable transfer times.

\subsubsection{Mars and Phobos}

For Mars, the injection from marginal Earth escape to Mars transfer
orbit requires a 0.4 km/s perigee burn and settling from the transfer
orbit to high Mars orbit another 0.7 km/s. Reaching low Mars orbit
needs additional impulsive 1.4 km/s or aerobraking. Thus, getting a
spacecraft to low Mars orbit with an E-sail would save 1.1 km/s of
impulsive \MARKI{$\Delta v$} and make aerobraking unnecessary.

If one wants to rendezvous with Phobos (6000 km orbital altitude),
aerobraking is less attractive because it would necessitate a 0.5 km/s
circularising burn to raise the perigee back up from the
atmosphere. Without aerobraking, Phobos rendezvous impulsive
\MARKI{$\Delta v$} distance from high Mars orbit is about 1 km/s.

A Phobos sample return mission therefore requires a total impulsive
\MARKI{$\Delta v$} of 2.1 km/s for landing \MARKIII{on} Phobos and another 1.7 km/s for the
return trip, if a high-speed hyperbolic Earth reentry is tolerated by
the sample capsule. In Phobos sample return, an E-sail would save 3.8
km/s of impulsive \MARKI{$\Delta v$} and enable a lighter Earth reentry shield
because reentry could occur from marginal Earth escape orbit without
hyperbolic excess speed. Saving 3.8 km/s of \MARKI{$\Delta v$} is equivalent to
saving about 2/3 of the initial mass if bipropellant hydrazine with
3.4 km/s specific impulse is employed. One also saves the mass of the
tanks and rocket engine, but on the other hand must include the
E-sail. Roughly speaking we expect these mass items to be of
comparable magnitude, so to a first approximation one can say that the
chemical propellant mass is saved.

One could also travel to Mars or Phobos by electric
propulsion. However, the total \MARKI{$\Delta v$} then becomes larger because the
Oberth effect is no longer utilised. The \MARKI{$\Delta v$} from marginal Earth
escape to high Mars orbit is then equal to the Earth-Mars orbital
speed difference 5.7 km/s and rendezvous with Phobos needs a further
2.1 km/s. The total \MARKI{$\Delta v$} for a Phobos sample return using electric
propulsion is then 15.6 km/s. For example with a typical 3000 s
specific impulse Hall thruster the xenon propellant fraction would
then be 40\%. Including the mass of the required high power electric power
system and solar panels would probably increase the initial mass near
the chemical propulsion value. Indeed, the Russian Phobos-Grunt
sample return mission which recently failed in Earth orbit did not use electric
propulsion but hydrazine.

The E-sail must produce the same total \MARKI{$\Delta v$} as electric
propulsion (or larger, because the E-sail thrust direction
controllability is more limited). However, being propellantless and
needing only modest electric power, the E-sail can deliver such high
\MARKI{$\Delta v$} without increasing the initial mass. The operability of the
E-sail in the vicinity of Mars has not been analysed in detail yet.
Expectedly the planet's plasma tail modifies the E-sail effect in the
nightside, but likely this does not change the above results
qualitatively. Because Mars does not have a strong intrinsic magnetic
field, its plasma environment is compact, not much wider than the
planet itself.

\subsubsection{Mercury}

Mercury is difficult to reach because it resides deep in Sun's gravity
well and because its low mass provides only a rather weak Oberth
effect to assist capture by impulsive propulsion. Mercury probes have
therefore made use of one or more gravity assist manoeuvres with
Venus, Earth and Mercury itself, resulting in long transfer
times. Despite its small dimension and long rotation period, Mercury
has a global, approximately dipolar magnetic field. The magnetopause
is located between 1000 and 2000 km from the planet's surface (on
average at 1.4 $R_{\rm M}$, but might even touch the planet when the
solar wind dynamic pressure is high). The strength of the field is
small compared to Earth and is $\sim 300$ nT at the equator.

\begin{figure}
\centerline{\includegraphics[width=1.65\columnwidth,angle=90]{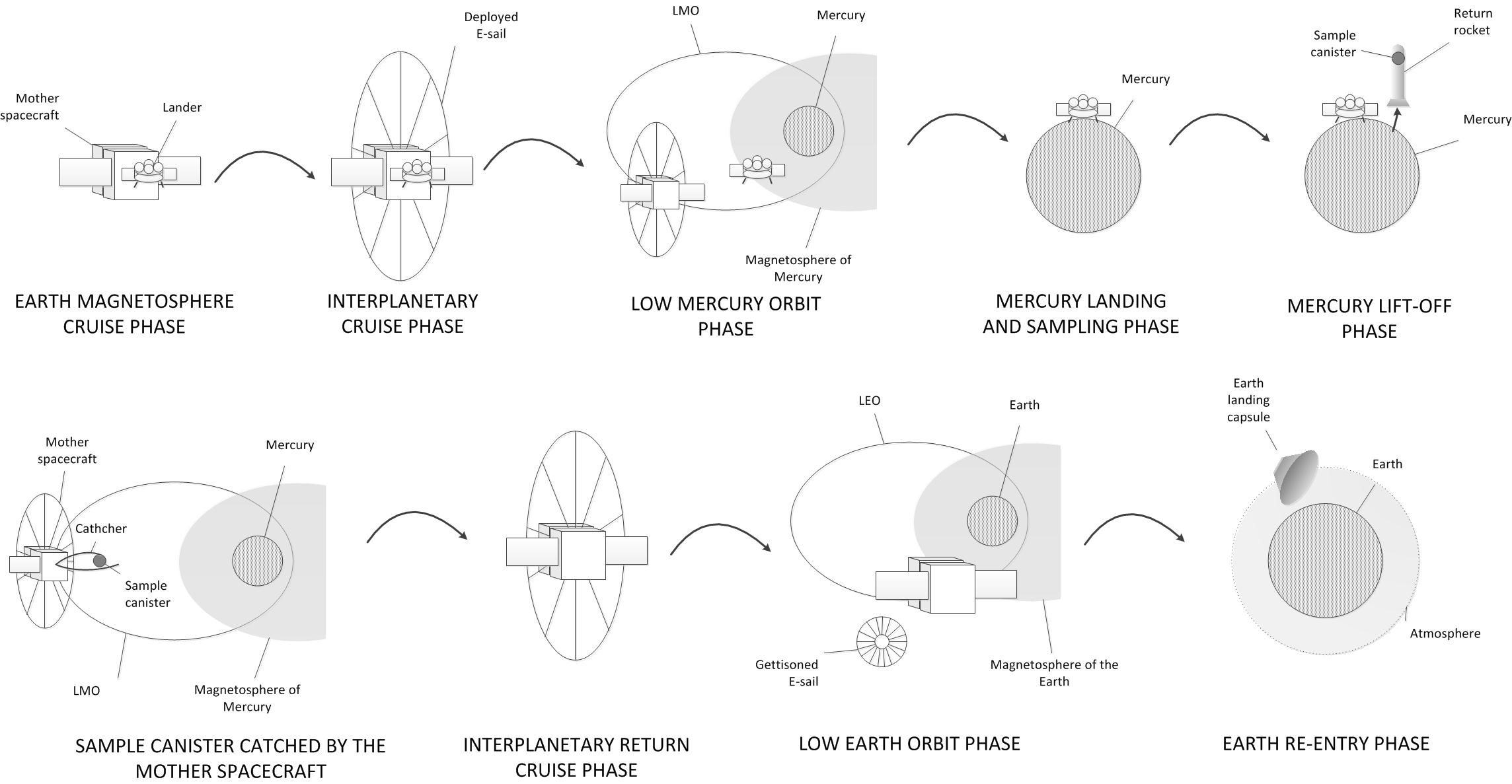}}
\caption{
Mercury sample return with single E-sail (see text). The E-sail in
relation to the main spacecraft is not drawn to scale.
}
\label{fig:Mercury}
\end{figure}

The newest mission under construction, BepiColombo, will spend more
than six years in transfer although it makes use of both chemical
propulsion and ion engines. In comparison, the E-sail could take a
probe to Mercury rendezvous (high planetary orbit) in less than one
year with no gravity assists \citep{QuartaEtAl2010}. Furthermore, it
could accomplish a return trip in the same time without extra
mass. Figure \ref{fig:Mercury} shows a scenario for returning a sample
from Mercury using a single E-sail in the following way: 1) The E-sail
flies to Mercury and settles to orbit the planet. 2) A lander
separates and lands on the surface by retrorockets. 3) The lander
picks up a sample and puts it in a small capsule which is mounted on a
small return rocket that was part of the lander payload. 4) The return
rocket lifts off to Mercury orbit where it jettisons the capsule which
also contains a radio beacon. 5) The E-sail mother spacecraft locates
the beacon signal and adjusts its orbit to rendezvous with the sample
capsule. In the approach phase cold gas thrusters are used for
fine-tuning the orbit. The capsule is picked up by a catcher and is
moved inside an Earth reentry shield inside the mother
spacecraft. Then the E-sail takes the mother spacecraft back to nearly
parabolic Earth orbit, the reentry capsule separates and returns to
Earth with the sample.

The presently developed aluminium E-sail tethers do not necessarily
tolerate the high temperature encountered on a Mercury mission unless
coated by a well emitting layer to assist cooling. Alternatively,
aluminium could be replaced by much more heat tolerant copper. A
similar ultrasonic bonding process to what we use with aluminium is
possible with copper as well.

\subsubsection{Asteroids}

Asteroids are targets where high \MARKI{$\Delta v$}'s produced by E-sail can be
uniquely useful because being lightweight targets, asteroids provide
essentially no Oberth effect to assist a capture by impulsive chemical
propulsion if a rendezvous is desired. Many asteroids are essentially
beyond reach for rendezvous by chemical propulsion and challenging to
reach by electric propulsion. Using E-sails to reach all potentially
hazardous asteroids was studied recently
\citep{QuartaAndMengali2010}. The E-sail would enable multi-asteroid
touring type missions where asteroids are studied in flyby and/or
rendezvous modes. With the propellantless E-sail, the only limit to
mission duration and the number of asteroids studied is set by the
durability of the equipment.

Asteroids are not only interesting scientifically, but also because of
the impact threat, asteroid resource utilisation and as potential
targets for manned exploration that need to be mapped beforehand.
The E-sail's very high lifetime-integrated total impulse per mass unit could
even be used to tow an Earth-threatening asteroid away from Earth's
path \citep{MerikallioAndJanhunen2010}. As an order of magnitude estimate, the
baseline 1 N E-sail could tug a 150 m asteroid away from Earth's path
in seven years \citep{MerikallioAndJanhunen2010}. There is clearly a
need for some further research to find out how to best attach the
E-sail to asteroids of different shapes and rotation states.

In addition to defleting a large asteroid, returning a small asteroid
to Earth orbit for scientific study has also been proposed using
electric propulsion \citep{BrophyAndGershman2011}.  In the mission
envisaged in \citep{BrophyAndGershman2011}, a solar electric propulsion
system is used to escape from Earth orbit and to rendezvous with a
near-Earth object (NEO). The first phase of the mission is then
devoted to the determination of the asteroid's trajectory and
spin. The spacecraft must reach the same rotation speed as the
asteroid, capture it, and let both masses slow down to rest. The
asteroid can then be moved to another trajectory or brought to Earth
orbit. The E-sail could be used for this kind of ``extended sample
return'' application as well, with the advantage of saving propellant
during the time needed to deflect the asteroid (depending on the
asteroid mass this period can last several years).


\subsection{Maintaining non-Keplerian orbits}

The propellantless thrust of an E-sail can be used to maintain a
non-Keplerian orbit, \MARKI{i.e.~an} orbit \MARKI{whose maintenance}
requires \MARKI{continuous propulsive} thrust
\citep{MengaliAndQuarta2009}. Such orbits enable a number of
qualitatively new applications, examples of which are discussed
next.

\subsubsection{Helioseismology from lifted orbit}

The spacecraft could orbit the Sun similarly to Earth, but in an orbit
which is lifted above the ecliptic plane \citep{MengaliAndQuarta2009}. From such orbit one would
have a continuous view to the polar region of the Sun, enabling e.g.~a
long time scale uninterrupted helioseismological coverage.

\subsubsection{Remote sensing of Earth and Earth's environment}

There are several conceivable E-sail orbits for remote sensing of the Earth
or the Earth environment. For example, so-called mini moons are small
asteroids that are temporarily captured by Earth's gravity field. The
mini moons are interesting e.g.~because they could be studied by a
varient of astronomical instruments from Earth at much closer range
than typical asteroids \citep{GranvikEtAl2012}. Mini moons typically enter the Earth system
near one of the Earth-Sun Lagrange points. Their entry could be
monitored by a spacecraft located on the sunward side of the Lagrange
L1 point, so that the mini moons would be visible to the spacecraft's
telescope at maximum solar illumination.

Several non-Keplerian orbits for Earth remote sensing, especially for
polar areas, have been investigated for solar photon sails and
electric propulsion, including figure eight shaped nightside orbits,
polar sitter orbits and non-Keplerian Molniya-type trajectories
\citep{CeriottiEtAl2012}. Many of the investigated orbits would also
suit the E-sail which can generally produce more thrust than other low
thrust methods. \MARKIII{The E-sail can provide higher thrust than an
  equal mass photonic sail.} The chief limitation of the E-sail in
this context is its inability to produce thrust inside the
magnetosphere where there is no solar wind.

\subsubsection{Giant planet auroras}

A planetary example of a non-Keplerian mission would be a spacecraft
which is lifted above the Jupiter-Sun Lagrange L1 point so that it has
a continuous view of Jupiter's north pole while being immersed in the
solar wind so that it can monitor it. The scientific application would
be to study to what extent Jupiter's auroras are driven by solar wind
changes. A traditional way of reaching this science goal would be to
have several spacecraft: one at Jupiter-Sun Lagrange point for
measuring the solar wind and additional ones in polar Jupiter orbit
for measuring the auroras continuously (since a single orbiter would
not be enough for continuous coverage of one of the
poles). Furthermore, in the traditional mission architecture the
orbiters would have to be radiation hardened to survive in the intense
Jovian radiation belts.

\subsubsection{Off-Lagrange point solar wind monitoring}

Nowadays, short-term forecasting of magnetospheric space weather
(magnetic storms) relies on continuously monitoring the solar wind
plasma density, plasma velocity and interplanetary magnetic field
(IMF) at the Earth-Sun Lagrange L1 point, by satellite such as ACE and
SOHO. Because it takes about one hour for the solar wind to travel
from the Lagrange L1 point to the magnetospheric nose, such monitoring
can give about one hour of warning time for preparing to the radiation
belt enhancement, geomagnetically induced current and other possible
adverse effects of magnetic storms. With its ability to generate
continuous thrust without consuming propellant, the E-sail could be used
to ``hang'' a spacecraft against the gravity field of the sun on the
sunward side of the Lagrange point, for example at twice the
distance from the Earth than the Lagrange point so that the warning
time would be two hours instead of one hour.

The high electric field of active E-sail tethers tends to disturb
plasma measurement and the current of the electron gun might
disturb the IMF measurement. In the context of solar wind monitoring,
many approaches are possible towards overcoming these problems. A
straightforward approach is to alternate the propulsive and
measurement phases in (say) 10 minute succession. This works if the
resulting data gaps are tolerated and if the E-sail can reach its full
thrust within the chosen gap duration. Another simple approach is to
use two identical spacecraft in nearby orbits which together can
produce a continuous realtime measurement of the solar wind. The third
option is to deploy the plasma measurement package with a
non-conducting $\sim 500$ m long tether from the E-sail mother
spacecraft so that the instruments are far from the influences of the
E-sail. Even when the E-sail is operating, the solar wind density can
be estimated by a simple omnidirectional onboard electron detector
which observes the thermal electron flux accelerated by the
voltage. The solar wind dynamic pressure might also be possible to
deduce from the produced thrust versus employed voltage, and the IMF
measurement might be possible despite the electron gun's influence
using a 5-10 m long fixed boom. In a full-scale mission the electron
gun current is maximally $\sim 50$ mA. By Biot-Savart law, the
magnetic perturbation of such current at 10 m distance is not more
than 1 nT which is smaller than a typical IMF at 1 au of $\sim 10$ nT.



\subsection{Near-sun missions}

The E-sail may be also successfully used to drive the spacecraft to
the vicinity of the sun. Depending on the target orbit, this type of
mission can be highly demanding in terms of \MARKI{$\Delta v$}, thus opening an
opportunity for the E-sail. As an example, for missions to near-sun
high inclination orbits, such as the Solar Orbiter (SOLO), the cost
related only to the orbit change manoeuvre would be at least 15 km/s
for reaching an inclination of 40$^{\rm o}$
\citep{SerafiniAndGonzalez}, a high price for either a chemical or
electric propulsion system.

When operating the E-sail towards the inner Solar System, the thrust
produced would increase inversely proportional to the distance from
the Sun, providing increased performances with respect to the nominal
E-sail operation at 1 au distance.  To realise this thrust increase,
the power consumption of the electron gun increases as the inverse
square of the distance i.e.~in the same way as the solar radiation
flux available to the solar panels. Since the efficiency of solar
panels typically gets degraded at high temperature, finding enough
electric power near the sun to run the E-sail at full voltage may
produce some technical challenges.

Due to the significant increase in temperature when cruising towards
the Sun, similar considerations on the material of the E-sail as those
made for the mission to Mercury may be applied (e.g., replacing
aluminium tethers by copper tethers having a significantly higher
tolerance for elevated temperature).

\subsection{Boosting to outer solar system}

The E-sail can be used as a ``booster'' for missions going to Jupiter
or beyond, and this can be done so that the E-sail operational phase
is limited to e.g. the 0.9-4 au solar distance range where the E-sail
hardware currently being prototyped is designed to operate. Thus, even
though the mission payload may require a nuclear power source if it
goes beyond Jupiter, the E-sail booster device can be solar powered
and not even need modifications of our present E-sail component
prototype specifications. Alternatively, a moderate extension of the
radial distance range to e.g. 0.9-8 au would increase the E-sail
performance in beyond-Saturn missions and still be solar powered. If a
nuclear power source is included onboard because of the needs of the
scientific payload, its power output could be used also by the E-sail
at large solar distances for somewhat increased
performance. Essentially, the E-sail provides a faster and lower mass
replacement for the inner planet gravity assist sequences which are
used by most present-day outer solar system missions. An additional
benefit of the E-sail is more frequent launch windows since planetary
gravity assist manoeuvres as typically not used.

Potential targets for outer solar system missions include the giant
planets Jupiter, Saturn, Uranus and Neptune and their moons, Jupiter
Trojan asteroids, Centaur objects, Kuiper belt objects and
beyond-heliosphere interstellar space. The giant planets provide large
enough Oberth effect to enable capture by modest-sized impulsive
chemical burn followed by E-sail boosting. The small outer solar
system objects such as Kuiper belt and Centaur objects provide no such
capability and for them the E-sail can feasibly provide only flyby
types of missions.

\begin{figure}
\centerline{\includegraphics[width=\columnwidth]{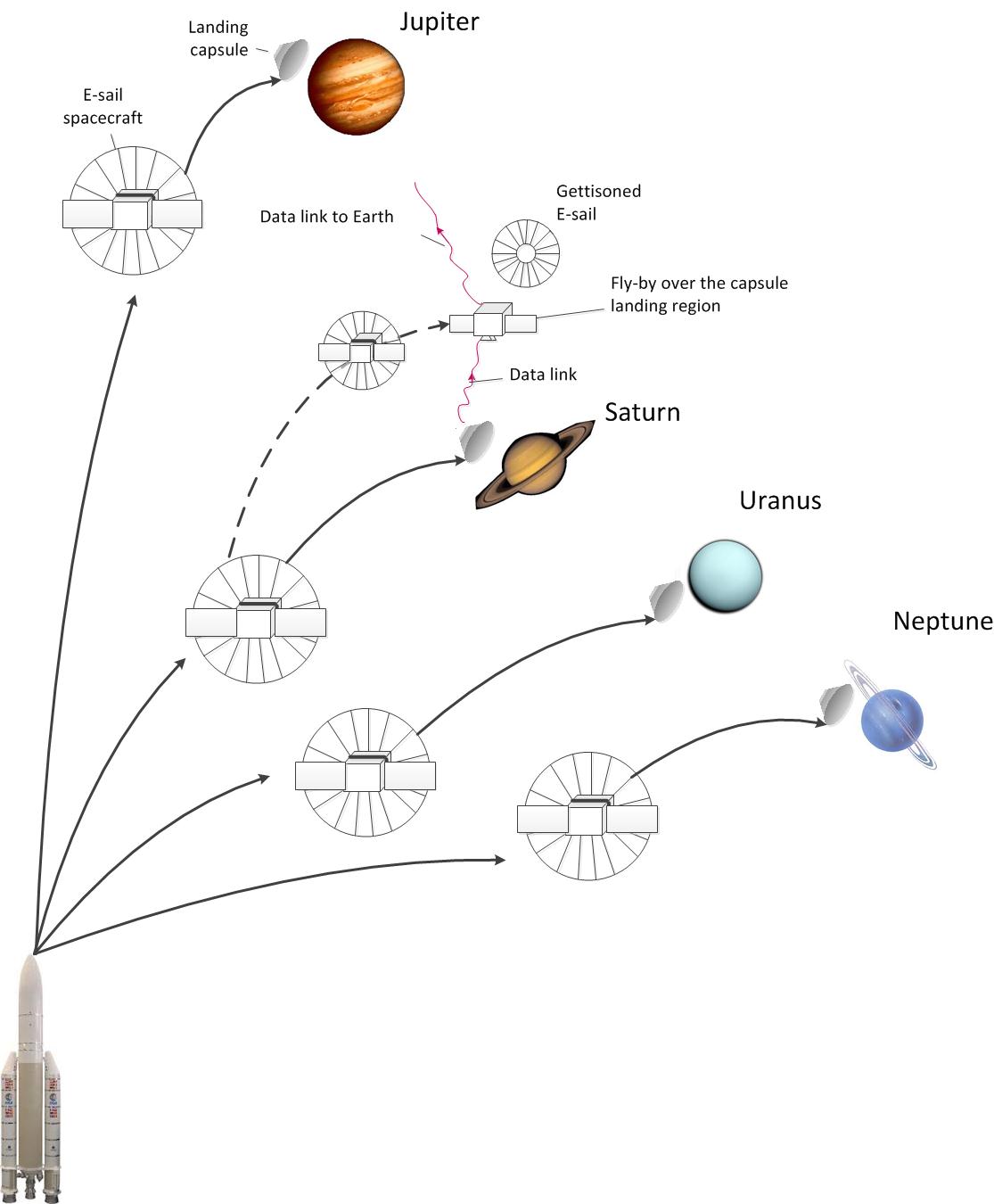}}
\caption{
Scenario for four identical atmospheric probes to all four giant
planets by a single launch. For easier visualisation, the details of
how the data from the atmospheric probe are relayed by the E-sail
mother spacecraft which flies by the giant planet are shown only in
Saturn's case. Illustrative photos courtesy by Arianespace, ESA and NASA.
}
\label{fig:Jupiter}
\end{figure}

For example, the isotope ratios of noble and other gases in the
atmospheres of the giant planets are well known only for Jupiter,
because of the successful measurements by the Galileo
entry probe. One could launch four E-sail equipped spacecraft, each
targeted to different giant planet, to release an entry probe near the
planet (Figure \ref{fig:Jupiter}). The entry probe would make a
high-speed atmospheric entry and then float downward, sending data to
the mother spacecraft which later sends it to the Earth. The Jupiter
Galileo probe already demonstrated successfully a high-speed entry;
other giant planets have smaller masses and therefore smaller entry
speeds. Because of the large mass of the giant planet, the relatively
high hyperbolic incoming speed of a fast E-sail probe (of order 15
km/s, for example) increases the atmospheric entry speed only
slightly. For decreased costs the probes sent to the four giant
planets might be identical. As with all E-sail spacecraft, any or all
of the probes can be launched with any launcher that reaches escape
orbit.

In one-way outer planet missions, the E-sail could act as a plug-in
replacement for present systems with no need to redesign the probe and
its payload, as long as the probe's mass is not too large (not more
than 1-1.5 tonnes, say) that it would be too heavy for an E-sail to
carry.

\section{Impactors and data clippers}

Without any onboard science instruments, an E-sailer could be used for
impacting a heavenly body at a given time, e.g.~when another
spacecraft is nearby and able to measure the properties of the impact
plume by remote sensing. For example, we could collide a small and
relatively fast E-sail with Jupiter's moon Europa, for example at the
same time when ESA's Jupiter Icy Moon Explorer (JUICE) makes its flyby of the moon around 2030. The high impact
speed should guarantee that the impactor's mass vaporises completely
so that there is no concern about contaminating Europa by
micro-organisms. The purpose of the E-sail impactor is to increase the
scientific output of another mission in a low-cost way.

The E-sail can accomplish a return trip with no propellant consumption
which is beneficial for sample return missions.  Within lower cost
mission category, the ability of the E-sail to return could be
utilised by ``data clippers'' \citep{Poncy2010}: spacecraft that carry
into Earth's vicinity large volumes of high resolution scientific data
stored in memory. Once in Earth's vicinity, the data can be
downloaded by low-cost ground antennas since the
available bandwidth scales as $\sim 1/R^2$ where $R$ is the data
transfer distance. The data clipper concept would enable a retrieval
of large amounts of data from distant targets by small missions whose
total cost has in principle no lower limit. The data clipper can
either collect the data by its own instruments or it can download at
the target from another perhaps more traditional spacecraft.

\section{Asteroid resource utilisation}

\begin{figure}
\centerline{\includegraphics[width=\columnwidth]{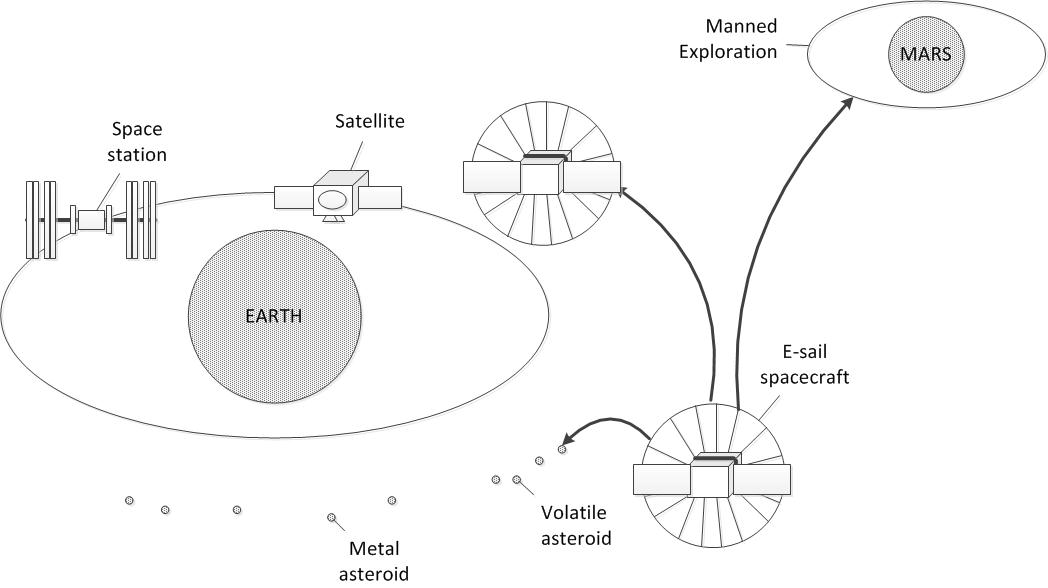}}
\caption{
Asteroid resource utilisation. Resources (for example water, other
volatiles, metals) are mined at asteroids and the materials are
transported by E-sails to serve lifting satellites to higher orbit and
manned exploration of Mars, asteroids and other bodies (LH2/LOX fuels,
oxygen, potable water).
}
\label{fig:ISRU}
\end{figure}

The ability of the E-sail to visit one or many asteroids repeatedly
and to carry payloads with very low propulsion system mass fraction
and zero propellant consumption would be well suited for mining water
and possibly other volatiles on asteroids, making rocket fuels such as
liquid hydrogen and oxygen out of them and transporting to orbits
where chemical fuels are needed (Figure \ref{fig:ISRU}). A chemical
rocket is the only known way of lifting a payload rapidly from LEO to
higher or escape orbit. Thus expectedly there should be a market for
asteroid-derived propellants in taking satellites to their orbit.

Asteroid-derived propellants could also facilitate manned Mars
exploration.  If the rocket is refuelled for the return trip in Mars
orbit, the propellant fraction is dramatically decreased and the
nuclear thermal rocket option becomes unnecessary. Besides for
propellant, manned exploration also needs water for drinking water and
breathing oxygen. Although one tends to circulate these substances,
the circulation may not be perfect.

Metal asteroids also contain significant concentrations of platinum
group metals. These metals may be precious enough to make it
economically feasible to mine them on asteroid and drop to Earth for
direct selling \citep{Gerlach2005}. Ultimately, given enough automation
and capital investments, a true space economy could arise where most
things in space are built mainly from asteroid or Moon derived raw
materials and not so much is needed to lift from Earth.

The E-sail can solve the asteroid transportation problem in an
economical and flexible way. Because no propellants are consumed by
E-sail transportation, also rocky and metal asteroids can be mined
directly without first having to transport the transportation
propellant from volatile-rich asteroids, for example.

\section{Challenging missions}

In general, the E-sail does not enable a speedy return from the outer
solar system because a pure E-sail return trip would use time which is
proportional to the orbiting period of the body from which the return
takes place because it depends on Sun's gravity to pull the spacecraft
inwards. The orbiting period increases as power 3/2 of the solar
distance and varies from 1.9 years for Mars to as much as 165 years
for Neptune. The exception is that from a nearly parabolic orbit
around one of the four giant planets one can inject the spacecraft at
significant speed towards the inner solar system by performing a
modest chemical burn near the planet and thus utilising the Oberth
effect. The E-sail can then be used in the inner solar system to
direct the probe into Earth rendezvous. Ideally, one would like to use
this approach both during the forward and return trips with the same
E-sail, but then the challenge is how to make the E-sail tether rig
survive the impulsive chemical burn without getting broken or tangled
up. A preliminary investigation of the parameters shows that survival
of the tether rig is probably possible, although one may need to
optimise the mass of the E-sail Remote Units or limit the heliocentric
travel speed or allow some increase in the chemical fuel
consumption. In principle it might also be possible to rewind the
tether rig back to their reels and then reopen them. Before there is
some practical experience of using the E-sail in space it may be difficult,
however, to assess if such backreeling procedure could be made
reliable enough. Consequently, we are currently not planning to make
any missions dependent on such reopening possibility.

If the tether rig survives the orbit insertion and de-insertion
chemical burns, this method allows one to get an E-sail spacecraft
into nearly parabolic giant planet orbit and back, using E-sail in
combination with a modest amount of chemical propulsion. To rendezvous
with a giant planet moon also requires a significant orbit change away
from parabolic and back, however. In principle it might be possible to
do this by E-sailing in the giant planet's corotating magnetospheric
plasma.  More detailed analyses are needed to investigate the
feasibility of E-sail based giant planet moon sample return missions.


\section{Conclusions}

We have given a brief survey of some of the E-sail missions that have
been thought of or analysed in more or less detail. Glossing over some
details, we summarise the results as follows. The E-sail can be used
to make most planetary missions cheaper or faster or both. It really
excels in asteroid missions and it makes two-way missions feasible
(sample return and data clippers). It can also be used as a booster
for outer solar system missions. The E-sail is enabling technology for
multi-asteroid touring and non-Keplerian orbit missions.

All E-sail missions must start from near or full escape orbit. Also,
any escape orbit is good for any E-sail probe. Hence, E-sails can be
piggybacked on other escape orbit launches and E-sails going to
different targets can be launched together in any combination.

In the longer run, E-sails might enable economic asteroid resource
utilisation and asteroid-derived propellant manufacture for satellite
orbit raising, for lifting E-sail missions themselves to escape
condition and for manned Mars and asteroid exploration.

{\it Acknowledgements\ }The research leading to these results has received funding from the
European Community's Seventh Framework Programme ([FP7/2007-2013])
under grant agreement number 262749. We also acknowledge the Academy
of Finland and the Magnus Ehrnrooth Foundation for financial support.

\clearpage


\end{document}